# Functional surfaces of laser-microstructured silicon coated with thermoresponsive PS/PNIPAM polymer blends: switching reversibly between hydrophilicity and hydrophobicity

Maria Kanidi[a,b], Aristeidis Papagiannopoulos[a], Andreea Matei[c], Maria Dinescu[c], Stergios Pispas[a], and Maria Kandyla[a, *]

[a]Theoretical and Physical Chemistry Institute, National Hellenic Research Foundation, 48 Vasileos Constantinou Avenue, 11635 Athens, Greece

[b]Department of Material Science, University of Patras, University Campus, 26504 Rio, Greece

[c]National Institute for Lasers, Plasma and Radiation Physics, 409 Atomistilor, 077125 Magurele-Bucharest, Romania

**Abstract**

We developed functional surfaces of laser-microstructured silicon coated with blends of polystyrene (PS) and poly(N-isopropylacrylamide) (PNIPAM) and we study their switching wetting behavior between hydrophilicity and hydrophobicity. Large areas of silicon are processed with reproducible surface micromorphology and spin-coated with PS/PNIPAM blends of two blend ratios. The wetting behavior of the surfaces is modified by the combination of surface topography and surface chemistry effects. PS/PNIPAM films are casted on flat and microstructured silicon substrates with or without a native $SiO_2$ layer. All films respond to the stimulus of temperature and films casted on microstructured silicon substrates with a native $SiO_2$ layer show the highest thermoresponsiveness presumably because they adopt a more favorable structure. Microstructuring provides a large specific area that extends the contact of PNIPAM chains with water molecules according to the Wenzel model, and thus increasing the film thermoresponsiveness, resulting in a reversible transition from hydrophilicity to hydrophobicity upon heating. The absence of the native $SiO_2$ layer from the silicon substrates affects the PS and PNIPAM arrangement in the films, increasing the water contact angle both below and above the lower critical solution temperature of PNIPAM and decreasing their thermoresponsiveness.

---

* Corresponding author: kandyla@eie.gr

## 1. Introduction

Control and switching of the wettability of a solid surface pose a challenge for several applications in the fields of industry and technology, such as waterproof textiles, coatings for boats, self-cleaning surfaces for solar energy panels, metal refining, microfluidics, as well as in the field of biology, such as drug delivery systems, cell encapsulation, immunoassays and biosensors, among others [1-5]. To this end, much attention is devoted to smart materials, which undergo switching of their wetting behavior between hydrophilicity and hydrophobicity by an external stimulus, such as temperature or pH treatment, light irradiation, electrical potential, mechanical manipulation, *etc.* [5]. Surface chemistry and topographic structure determine the surface wettability, which is measured by the water contact angle $\theta$ between the surface and water [6]. The wettability presents mainly two states, hydrophilicity ($\theta = 10°–90°$) and hydrophobicity ($\theta$ = 90°–150°), and their extreme counterparts, superhydrophilicity ($\theta$ = 0°–10°) and superhydrophobicity ($\theta$ = 150°–180°) [6]. Stimuli-responsive surfaces enable the reversible control of wettability. Such surfaces can be developed by either chemical modification with polymer films, brushes, and self-assembled monolayers [7], or micro/nanostructuring of their topography [8, 9], or the combination of both [10-12].

One way to modify the surface chemistry is the deposition/attachment of functional polymer films. A widely known stimuli responsive polymer is poly(N-isopropylacrylamide) (PNIPAM), which exhibits a reversible phase transition at the lower critical solution temperature (LCST) of 32°C. Below the LCST, the PNIPAM chains are able to form intermolecular hydrogen bonds with water molecules in aqueous solutions, displaying hydrophilicity. Heating above the LCST is accompanied by a compact and collapsed conformational change in the polymer chains, obstructing the interaction of water with the hydrophilic groups C=O and N–H of PNIPAM and resulting in a less hydrophilic behavior of the material [13]. Because switching of PNIPAM wettability occurs at 32°C, which is close to human body temperature, PNIPAM in hydrogel systems is widely used for biological applications [14]. Additionally, PNIPAM is used for the development of stimuli responsive surfaces with a controllable wetting behavior via grafting techniques. Specifically, PNIPAM is grafted on surfaces alone or in combination with other polymers (polystyrene (PS), poly(methylmethacrylate) (PMMA), poly(N-dimethylacrylamide) (PDMA), pentamethyl diethylenetriamine (PMDETA), and polyacrylic acid (PAA)) via reactions of free radical polymerization, surface-initiated atom transfer radical polymerization (SI-ATRP), and plasma-induced grafting polymerization [3, 13, 15-20]. A significant advantage of grafting techniques is that they offer precise localization of the polymeric chain on the surface and accurate control of the chain arrangement within the polymeric film [21]. Additionally, grafting techniques allow the formation of stable layers of (co)polymers with various compositions, functionalities, and controlled topography [21]. On the other hand, such techniques require elaborate equipment and metal catalysts, constituting costly and complicated approaches, partially due to the removal of catalysts at the end of the polymerization process.

Besides the chemical composition of a surface, another factor that affects wettability is surface topography. The angle of contact between water and a surface is affected by inducing roughness on the surface [22]. Two models, Wenzel and Cassie-Baxter, describe the wetting behavior of a liquid on a rough surface [23]. According to the Wenzel model, the liquid conforms to the topography of the surface and the surface roughness can enhance both hydrophilicity and hydrophobicity. Specifically, when roughness is induced, a hydrophilic surface can become more hydrophilic and a hydrophobic surface can become more hydrophobic [24]. According to the Cassie-Baxter model, the liquid conforms only to the top of corrugations due to air trapping underneath, thus roughness contributes significantly to hydrophobicity enhancement [23, 24]. The model that describes the wetting behavior of a rough surface, and thus the increase or decrease of contact angles, depends strongly on the corrugation shape and size [22]. The synergy of the surface chemical composition and topography is critical to generate functional surfaces with tailored wettability, inducing advanced properties, such as water repellency, anti-adhesion, anisotropic dewetting, *etc.* [1]. To modify the surface topography and induce micro- and nanopatterning, there are several methods, such as lithography [25], electrochemistry [26], plasma and wet etching [27, 28], among others. Laser processing is a complementary, cost-effective, and simple method for tailoring micro- and nanostructures over large areas on metals, semiconductors, glasses, polymers, and other materials [29-31].

In this work, we develop functional surfaces of laser-microstructured silicon coated with PS/PNIPAM polymer blends and we study their switching wetting behavior between hydrophilicity and hydrophobicity upon heating. Based on our previous work on the wettability of PS/PNIPAM blend systems casted on flat silicon substrates, which showed a tunable, thermoresponsive wetting behavior [32], we take advantage of the large specific area of microstructured silicon to enhance the PS/PNIPAM film thermoresponsiveness. Silicon surfaces are microstructured via laser processing, inducing roughness on the surface, and spin-coated with PS/PNIPAM thin films of two blend ratios. Blending the thermoresponsive PNIPAM with PS provides mechanical stability to the films, because PS is in the solid state at room temperature since it has a high glass transition temperature ($T_g$ ~100 $^o$C), above which it flows, which makes PS suitable for high-quality surface coatings. In order to study the combined effect of micromorphology and surface chemistry of the silicon substrate, PS/PNIPAM films are casted on flat and microstructured silicon substrates with or without a native $SiO_2$ layer. All PS/PNIPAM films respond to the stimulus of temperature, however the degree of response depends on the underlying substrate. Films casted onto microstructured silicon substrates demonstrate increased water contact angles, compared to films on flat silicon, already at room temperature and switch reversibly from hydrophilic to hydrophobic upon heating. On the other hand, even though films on flat silicon also show an increase in the water contact angle upon heating, they remain hydrophilic and do not demonstrate a transition of their wetting state. The absence of the native $SiO_2$ layer from silicon substrates results in increased water contact angles of PS/PNIPAM films and decreased thermoresponsiveness. Laser processing is a single-step, maskless, tabletop method that allows quasi-uniform and controllable patterning of silicon surfaces [33, 34]. Additionally, spin coating is a simple and reproducible technique for film development, offering control of the process and avoiding the drawbacks of grafting techniques, such as the use of elaborate equipment [21, 35]. The combination of thermoresponsive polymer systems with reversibly switchable wettability and micro-patterned surfaces paves the way for the development of smart surfaces, able to extend and enhance traditional wetting applications by simple and cost-effective processes.

## 2. Materials and Methods

2.1 Materials. Polystyrene ($PS_{1488}$) homopolymer ($M_w$=155,000, $M_w/M_n$=1.05) was synthesized by anionic polymerization and poly(N-isopropylacrylamide) ($PNIPAM_{265}$) homopolymer ($M_w$=30,000, $M_w/M_n$=1.16) was synthesized by RAFT polymerization. THF solvent of analytical grade (Sigma-Aldrich) was used without further purification. For contact angle measurements, we used freshly distilled water from an all-glass distillation apparatus.

2.2 Microstructured silicon substrates. Silicon wafers ((100), thickness $500 \pm 25$ μm) were cleaned in an ultrasonic bath of acetone and methanol for 15 min before laser microstructuring. Microstructured silicon substrates were fabricated by nanosecond laser processing in a gas environment of sulfur hexafluoride ($SF_6$). A pulsed Q-switched

Nd:YAG laser system was used with 1064 nm wavelength, 5 ns pulse duration, and 10 Hz repetition rate. A silicon wafer was placed in a vacuum chamber, which was filled with 0.6 bar $SF_6$. The laser beam was focused on the silicon surface through a quartz window using a lens of 200 mm focal length and the vacuum chamber was raster scanned using a computer-controlled set of *xy* translation stages. The scanning speed was set so that each spot on the silicon surface was irradiated by 1000 laser pulses. The average fluence was ~1 $J/cm^2$ on the silicon surface. For the removal of the native silicon oxide ($SiO_2$) layer from silicon surfaces, flat and microstructured silicon substrates were immersed for 5 minutes in an aqueous solution of 5% hydrofluoric acid (HF). After HF etching, silicon surfaces were cleaned with distilled water and methanol and were dried in nitrogen ($N_2$) gas flow.

2.3 Polymer film preparation. Blends of $PS_{1488}$/$PNIPAM_{265}$ were prepared in 75/25 and 50/50 blend weight ratios dissolved in THF, consisting of 5 wt% polymer solution, and stayed overnight. Blend films were spin-casted onto clean flat and microstructured silicon substrates, with or without a native $SiO_2$ layer, at 3000 rpm for 30 sec. After spin coating, the films were dried in ambient conditions at room temperature for one hour. In order to optimize the film quality, we tried other blend ratios as well. PS/PNIPAM films with a lower PNIPAM ratio did not present repeatable thermoresponsivity. On the other hand, films with a higher PNIPAM ratio were not mechanically stable, due to the water solubility of PNIPAM. The PS/PNIPAM ratio should be adjusted to provide films with mechanical stability, induced by PS, and a repeatable wetting behavior. Therefore, we choose the 75/25 and 50/50 blend ratios in order to obtain high-quality films with reliable wettability. All films were characterized by scanning electron microscopy (SEM) using a field emission microscope, energy dispersive spectroscopy (EDS), micro-Raman spectroscopy, and X-ray photoelectron spectroscopy (XPS). Also, the film thickness on flat silicon substrates was measured by profilometry and films of 75/25 and 50/50 PS/PNIPAM are 700 – 800 and 500 – 600 nm thick, respectively.

2.4 Micro-Raman spectroscopy. Raman spectra were acquired with a Renishaw inVia Reflex Raman microscope, equipped with a Peltier-cooled charge coupled device (CCD) and a motorized *xyz* microscope stage with a lens of magnification ×100, in a backscattering geometry. The 514.5 nm line of an argon laser was used for excitation. Together with the rest of the system configuration (grating, slit width, CCD partition) this results in a spectral resolution of ~1 cm$^{-1}$. The laser beam was focused on the sample to a spot diameter 1–2 μm and the excitation laser power was 0.10 mW.

2.5 X-ray photoelectron spectroscopy (XPS). X-ray photoelectron spectroscopy was performed using an ESCALAB™ XI+ spectrometer (Thermo Scientific, USA) with a monochromatic Al Kα source at 1486.6 eV. The XPS survey spectra were recorded with a step size of 1 eV and a pass energy of 50 eV, as a result of 5 scans. The high-resolution spectra for the $C_{1s}$ were recorded with 0.1 eV energy step size and 10.0 eV pass energy. For $C_{1s}$, 10 scans were acquired. For depth profile analysis that we used in order to investigate the ratio PNIPAM/PS towards the bottom of the film, between 20 and 25 layers were analyzed, the etching was made using a 300 atoms cluster gun, working at 2000 eV. Each level was etched for 300 seconds.

2.6 Contact angle measurements. A water drop of 5 μL was deposited on the surface of each film to measure the contact angle at room temperature (25$^{o}$C) and at 45$^{o}$C, using a constant temperature control base. Images of water drops on the surface were obtained by a 2.0MP 500x USB Digital Microscope and analyzed by standard MATLAB functions. In detail, the images were transformed to grayscale and subsequently the drop edge profile was determined. The profile was numerically analyzed in polar coordinates ($r$, $\theta$). The origin of the coordinates was chosen on the intersection between the apparent perpendicular axis of symmetry of the drop and the solid/liquid interface. The profiles $r(\theta)$ obtained this way vary smoothly with $\theta$ and were easily modeled by fourth-order polynomials. The contact angles were extracted from the derivatives (transformed back to Cartesian coordinates) on the left and right-side contacts of the drop with the surface. Typically, the deviation between left and right contact angles was less than 2°. The average of the two values is reported as the measured contact angle. For each film, water contact angles were measured three times.

## 3. Results and Discussion

SEM micrographs of microstructured silicon surfaces with a native $SiO_2$ layer and without it are presented in Figs. 1a and b, respectively. In both Figs., the silicon surface is covered by a quasi-ordered array of conical microstructures with a mean half-height diameter of 15 ± 2 μm, a mean height of 65 ± 5 μm, and a mean distance between neighboring microstructures 16 ± 3 μm center-to-center. Comparing Fig. 1a with Fig. 1b, we observe that the removal of the native $SiO_2$ layer does not affect the morphology of silicon microstructures. Using laser microstructuring, we are able to process large areas of silicon with quasi-uniform micromorphology. The formation of such microstructures is the result of melting and interference effects that occur upon nanosecond laser irradiation of silicon in $SF_6$ environment [36, 37]. Laser processing allows the control of surface morphology by tuning the fabrication parameters, such as wavelength, pulse duration, the gas or liquid environment, pressure of gas environment, laser fluence ($J/cm^2$), and number of incident laser pulses [30, 38, 39].

After laser processing, microstructured silicon substrates are coated with PS/PNIPAM blends of 75/25 and 50/50 blend ratios. In Figs. 2a and b, SEM micrographs show a 50/50 PS/PNIPAM film casted onto a microstructured silicon substrate, filling the valleys between microstructures. Figure 2b shows a film area of high magnification where it is not clear whether the film covers entirely the microstructures or it is only concentrated in the valleys between them. A layered image from EDS elemental analysis shows the spatial distribution of the chemical elements that constitute the surface (Fig. 3a). Mapping analysis shows the detected elements (silicon, carbon, oxygen, and nitrogen) and the distribution of silicon, carbon, and nitrogen on the surface (Fig. 3b). The elements of carbon and nitrogen correspond to the PS/PNIPAM film coating and they are detected spatially on the entire surface, verifying that the film covers the silicon microstructures, as well as the valleys between them. Similar results were observed for a 75/25 PS/PNIPAM film casted onto a microstructured silicon substrate (not shown here).

To investigate the topography of the PS/PNIPAM blend films we perform micro-Raman spectroscopy on films casted on microstructured silicon substrates at ~25 $^o$C (below the LCST of PNIPAM). For both blend ratios, Raman spectra on three different locations of the same film show peaks that correspond to the vibrational modes of C-H for PS and PNIPAM chains in the spectral range 2800–3200 $cm^{-1}$ (Figs. 4a and b). The vibrational mode of C-H for PS chains appears at 2852 $cm^{-1}$, 2907 $cm^{-1}$, 2976 $cm^{-1}$, 2999 $cm^{-1}$, and 3054 $cm^{-1}$, and the vibrational mode of C-H for PNIPAM chains appears at 2874 cm$^{-1}$, 2921 cm$^{-1}$, 2935 cm$^{-1}$, and 2976 cm$^{-1}$, verified by Raman spectroscopy on solid PS and PNIPAM [32]. Previous studies have shown these PNIPAM Raman peaks down-shift to lower wavenumbers when heating above the PNIPAM LCST, as a result of the PNIPAM phase transition to a compact and collapsed conformational state, which also induces the hydrophobic behavior of the material [40-42]. Comparing the three Raman spectra shown in Fig. 4a, spectrum # 1 presents peaks that correspond only to PS, indicating the absence of PNIPAM in this area, in contrast with spectra #2 and #3 that present peaks that correspond to both polymers. The absence of PNIPAM peaks in spectrum #1 is due to macrophase separation, that occurs between the components of the homopolymer blend [32, 43]. Additionally, the low ratio of PNIPAM in the blend of Fig. 4a (25%) results in the detection of low intensity PNIPAM Raman peaks compared to the intensity of PS Raman peaks. In case of the 50/50 PS/PNIPAM film (Fig. 4b), spectrum #1 presents peaks that correspond only to PS, spectrum #2 presents peaks that correspond to both PS and PNIPAM, and spectrum #3 presents peaks that correspond only to PNIPAM. Similar to the 75/25 PS/PNIPAM film, the 50/50 PS/PNIPAM film presents macrophase separation and the high PNIPAM ratio in the blend (50%) allows the detection of areas corresponding to each homopolymer separately in addition to areas corresponding to both. Phase separation can determine the surface topography, due to the formation of micro- and nano-domains on the film surface, creating a complex morphology [32]. In case of blend films casted on microstructured silicon substrates, we do not observe micro- and nano-domains on the film surface in Figs. 2a and b due to the micromorphology of the substrate.

Water contact angle measurements of PS/PNIPAM films casted on flat and microstructured silicon surfaces, with or without a native $SiO_2$ layer, are presented in Fig. 5. All water contact angles are measured at room temperature (~25$^o$C) in order to study the effect of the substrate micromorphology and chemistry on the wetting behavior of PS/PNIPAM films. For reference, we also measured water contact angles on bare flat and microstructured silicon surfaces, with or without a native $SiO_2$ layer. Both flat and microstructured silicon surfaces with a native $SiO_2$ layer present a similar hydrophilic wetting behavior with a water contact angle of ~55$^o$. However, the removal of the native $SiO_2$ layer affects significantly their wetting behavior. Specifically, the flat silicon surface shows an increase in water contact angle by ~26$^o$ upon the removal of the native $SiO_2$ layer, still displaying hydrophilicity albeit decreased, while the microstructured silicon surface shows an increase in water contact angle by ~67$^o$ upon the removal of the native $SiO_2$ layer, which renders it hydrophobic. The removal of the native $SiO_2$ layer from the microstructured silicon surface reveals the effect of micromorphology on its wetting behavior, enhancing the water contact angle due to air trapping between the rough surface and water [23]. Films of 75/25 and 50/50 PS/PNIPAM blends casted on flat silicon substrates without a native $SiO_2$ layer present an increase in their water contact angle by ~3$^o$ and ~9$^o$, respectively, compared to that with a native $SiO_2$ layer. Similarly, films of 75/25 and 50/50 PS/PNIPAM blends casted on microstructured silicon substrates without a native $SiO_2$ layer show an increase in their water contact angle by ~20$^o$ and ~15$^o$, respectively, compared to that with a native $SiO_2$ layer. As we show below, the absence of the native $SiO_2$ layer from silicon substrates results in PS enrichment of the top part of PS/PNIPAM films, which is hydrophobic, increasing their water contact angle. Comparing the wetting behavior of 75/25 and 50/50 PS/PNIPAM films casted on flat silicon substrates with or without a native $SiO_2$ layer, we observe that films with a higher PS ratio (75%) present higher contact angles than films with lower PS ratio (50%), because PS is hydrophobic with a water contact angle of ~100$^o$ [32]. On the other hand, we observe that films of both blend ratios, casted on microstructured silicon substrates with or without a native $SiO_2$ layer, present a similar wetting behavior regardless of the blend ratio. In the case of films casted on flat silicon substrates, the wetting behavior is determined by the chemical composition of the film, while the wetting behavior of the films casted on microstructured silicon substrates is determined by the substrate micromorphology.

Figure 6a shows water contact angle measurements on PS/PNIPAM films of both blend ratios casted on flat and microstructured silicon substrates with a native $SiO_2$ layer at 25$^o$C and 45$^o$C. Upon heating, films of 75/25 blend ratio show an increase in their water contact angle by ~15$^o$ (on flat silicon) and ~22$^o$ (on microstructured silicon). An increase in the water contact angle by ~18$^o$ and ~34$^o$ between 25$^o$C and 45$^o$C is observed for films of 50/50 blend ratio casted on flat and microstructured silicon substrates, respectively, which is higher compared to the increase of films of 75/25 blend ratio. All PS/PNIPAM films casted on flat and microstructured silicon substrates respond to the stimulus of temperature and their thermoresponsiveness is correlated with the blend ratio. Specifically, films with a higher ratio of PNIPAM (50%) present higher responsiveness compared to those with a lower ratio of PNIPAM (25%). Films casted on microstructured silicon substrates present such a high increase of water contact angle upon heating, that their wetting behavior switches from hydrophilic to hydrophobic. Microstructuring of silicon substrates enhances the film thermoresponsiveness due to their large specific area. Indeed, the micromorphology of these silicon substrates allows extended contact between water and the PS/PNIPAM film surface, which contains the thermoresponsive PNIPAM chains. The wetting behavior of PS/PNIPAM films casted on microstructured silicon substrates is described by the Wenzel model since films show high thermoresponsiveness, which is attributed to the effective interaction between water and PNIPAM chains as the water droplet conforms to the surface topography.

Figure 6b shows the effect of the absence of $SiO_2$ from the silicon substrates on the wetting behavior of 75/25 and 50/50 PS/PNIPAM films upon heating. All films preserve their thermoresponsiveness, depending on the percentage of PNIPAM in the blend ratio, and the wetting behavior of the films casted on microstructured silicon substrates switches, becoming hydrophobic upon heating, similar to the films casted on substrates with a native $SiO_2$ layer (Fig. 6a). In Fig. 6b, we observe that films casted on microstructured silicon substrates present higher thermoresponsiveness than those casted on flat silicon substrates, due to the extended contact between water and PNIPAM chains, provided by the large specific area of the microstructures, according to the Wenzel model. Comparing the water contact angle measurements of Fig. 6a with Fig. 6b, the removal of the native $SiO_2$ layer from silicon substrates results in higher water contact angles of PS/PNIPAM films both below and above 32$^o$C. Also, the thermoresponsiveness of films casted on silicon substrates with a native $SiO_2$ layer is higher than that of films casted on substrates without a native $SiO_2$ layer. We attribute this fact to the arrangement of PS and PNIPAM in the film, depending on the presence or absence of the native $SiO_2$ layer on the underlying silicon substrate, as shown below by XPS measurements.

Figure 7a shows an image of a water droplet on a 50/50 PS/PNIPAM film casted on microstructured silicon with a native $SiO_2$ layer below and above $32^oC$. Upon heating, the 50/50 PS/PNIPAM film shows an increase in water contact angle by $\sim34^o$ and switches from hydrophilic to hydrophobic, presenting the highest thermoresponsiveness of all films. Additionally, Figure 7b shows the reversible switching behavior of the 50/50 PS/PNIPAM film on microstructured silicon, demonstrating the transition between hydrophilicity and hydrophobicity for seven cycles of heating/cooling.

To better understand the effect of surface chemistry of the silicon substrate on the arrangement of PS and PNIPAM in the films and hence their wetting behavior, we used X-ray photoelectron spectroscopy. The elemental quantification (%) of 50/50 PS/PNIPAM films, casted on flat silicon substrates with or without a native $SiO_2$ layer, is presented as a function of etching layers in Fig. 8. An example of an XPS depth profile survey is presented in supplementary material, Fig. S1. The percentage of carbon is attributed to both polymers of the blend film, the percentage of nitrogen is attributed to PNIPAM, and the percentage of oxygen is attributed to PNIPAM and the native $SiO_2$ layer on the substrate, thus the percentage of oxygen is lower in case of the film casted on silicon without a native $SiO_2$ layer (Fig. 8b). In Fig. 8a, we observe that the percentage of nitrogen is higher in the upper etching layers and lower in the bottom film layers, for films casted on silicon with native $SiO_2$. On the other hand, in Fig. 8b we observe that the percentage of nitrogen remains constant regardless of the etching layer. These results indicate that in the presence of native $SiO_2$, PNIPAM is distributed mainly in the upper film layers, while in the absence of native $SiO_2$, PNIPAM is distributed uniformly throughout the film.

Figure 9 shows the variation of the relative amount of N-C=O and the shake-up $\pi$-$\pi$* transition populations attributed to PNIPAM and PS, respectively, in a 50/50 PS/PNIPAM film casted on a flat silicon substrate with a native $SiO_2$ layer (Figs. 9a and b) and without it (Figs. 9c and d), as a function of etching layer number, calculated by the area fitting of the C1s peak from spectra acquired by XPS. We describe how we calculate the ratios of PNIPAM and PS in supplementary material, Fig. S2 and Table S1. In the case of the silicon substrate with native $SiO_2$, we observe higher values of PNIPAM ratio in the upper film layers, due to more effective enrichment of PNIPAM in the upper part of the film (interfacing the air), compared to the bottom part of the film (interfacing $SiO_2$) (Fig. 9a). In Fig. 9b, we observe higher values of PS ratio in the bottom film layers, close to the silicon/$SiO_2$ substrate, resulting in more effective enrichment of PS in the bottom part of the film. Comparing Fig. 9a with Fig. 9c, we observe that the decrease rate of the PNIPAM ratio towards the bottom of the film is higher in the case of the film casted on silicon with $SiO_2$ than without $SiO_2$, as shown by the slope of the fitting lines. Furthermore, the slope of the fitting line of the PS ratio is higher in Fig. 9b than in Fig. 9d, indicating that the increase rate of the PS ratio towards the bottom of the film is higher in case of the film casted on silicon with $SiO_2$ than without $SiO_2$. Therefore, in the presence of native $SiO_2$, the bottom part of the films is more enriched with PS than in the absence of native $SiO_2$. On the other hand, PNIPAM is arranged preferentially in the top part of the films in the presence of native $SiO_2$, while the absence of native $SiO_2$ results in a decrease of PNIPAM in the upper part of the films. Consequently, the higher relative amount of PNIPAM in the upper part of the films on silicon with $SiO_2$ results in higher thermoresponsiveness of these films, compared to those casted on silicon without $SiO_2$, as shown in Fig. 6.

In summary, all PS/PNIPAM films, casted on microstructured and flat silicon substrates with or without a native $SiO_2$ layer, respond to the stimulus of temperature. The observed thermoresponsiveness is attributed to the presence of PNIPAM and films with a higher PNIPAM ratio present higher thermoresponsiveness, regardless of the morphology and surface chemistry of the silicon substrate. The removal of $SiO_2$ from the silicon substrates, either flat or microstructured, increases the room temperature water contact angle of the overlying films but does not enhance their thermoresponsiveness. On the contrary, the presence of the native $SiO_2$ layer on silicon results in PNIPAM enrichment of the top part of the films (interfacing with air and the water droplet), resulting in higher film thermoresponsiveness. In the presence of the native $SiO_2$ layer, we observe that films casted on microstructured silicon substrates exhibit high thermoresponsiveness, measuring up to ~$34^o$ increase in water contact angle upon heating, and switch reversibly between hydrophilicity and hydrophobicity, while films casted on flat silicon substrates do not undergo switching of their wetting behavior, even though they respond to the stimulus of temperature, measuring up to ~$18^o$ increase in water contact angle upon heating. The latter case agrees with our previous study on spin-casted PS/PNIPAM blend films on flat silicon [32] and with other studies, which graft PNIPAM on flat silicon surfaces, which showed an increase of the water contact angle upon film heating without switching between hydrophilicity and hydrophobicity [3, 13, 16, 17, 19, 20]. However, grafted PNIPAM films on non-planar substrates, such as silicon nanowires and microgrooves, show a transition of their wetting behavior by varying the micro/nanostructure size and spacing [11, 13]. Microstructuring induces roughness on the surface, which constitutes a crucial parameter for wettability, depending on the morphological characteristics of the microstructure [13, 22, 44]. Using grafting techniques, a PS layer can be grafted on a silicon substrate and a PNIPAM layer can be grafted above the PS layer on the upper part of the film. Grafting PNIPAM preferably on the upper part of the film, the PNIPAM chains are localized on the film-air interface, being able to interact with water molecules and respond effectively upon heating. Nevertheless, chain grafting can induce constrains to polymer chain rearrangement/relocalization and conformational changes, which may in turn hinder the response of the formed film to the external stimulus of temperature. In contrast to grafting techniques, the polymer arrangement in blend films deposited by spin coating cannot be controlled, because polymers adopt their thermodynamically favored structure. In this work, we demonstrate that spin-casted PS/PNIPAM films on laser-microstructured silicon substrates with native $SiO_2$ adopt a favorable structure and present significantly enhanced reversible thermoresponsiveness, which switches their wetting behavior between hydrophilicity and hydrophobicity.

## 4. Conclusions

We developed laser-microstructured silicon substrates coated with PS/PNIPAM blend films and we studied their thermoresponsive wetting behavior as a function of substrate micromorphology and surface chemistry. Taking advantage of the scalability that the laser patterning method offers, we can process large areas of silicon with reproducible surface micromorphology in order to use them as substrates for film casting, inducing 3D topography to the casted films and tuning their wetting behavior. PS/PNIPAM films of 75/25 and 50/50 blend ratios were spin-casted on microstructured silicon substrates with or without a native $SiO_2$ layer. We take advantage of the large specific area of the silicon substrates in order to enhance the film thermoresponsiveness. All films show an increase in water contact angle as they respond to the stimulus of temperature, but films with a higher ratio of PNIPAM exhibit higher thermoresponsiveness. The absence of the native $SiO_2$ layer results in higher water contact angles at $25^{o}C$ and $45^{o}C$ but the PS/PNIPAM films present lower thermoresponsiveness, compared to the films casted on silicon substrates with a native $SiO_2$ layer. This is because the presence of the native $SiO_2$ layer results in PNIPAM enrichment in the top part of the film, which comes into contact with water. Comparing the wetting behavior of films casted on flat and microstructured silicon substrates, we note that microstructuring provides a large specific area that extends the contact of PNIPAM chains with water molecules according to the Wenzel model. This enhances the film thermoresponsiveness to achieve switching from hydrophilicity to hydrophobicity. We also demonstrate that the transition is reversible and lasts for several heating/cooling cycles. Using simple, rapid, and cost-effective techniques, such as spin coating and laser processing, we combine thermoresponsive polymer systems with micro-patterned substrates for the development of smart functional surfaces with reversibly switchable wettability, able to extend and improve traditional wetting applications.

**Acknowledgements**

We acknowledge support of this work by the project "Advanced Materials and Devices" (MIS 5002409) which is implemented under the "Action for the Strategic Development on the Research and Technological Sector", funded by the Operational Programme "Competitiveness, Entrepreneurship and Innovation" (NSRF 2014-2020) and co-financed by Greece and the European Union (European Regional Development Fund). M. Kanidi acknowledges support through a Ph.D. fellowship by the General Secretariat for Research and Technology (GSRT) and the Hellenic Foundation for Research and Innovation (HFRI). We also thank Dr. C. Chochos for his help with SEM and EDS measurements.

**References**

1. M. Ma, M. R. Hill, Superhydrophobic surfaces, Current Opinion in Colloid & Interface Science, 11, 2006, 193-202.

2. F. Xia, Y. Zhu, L. Feng, L. Jiang, Smart responsive surfaces switching reversibly between super-hydrophobicity and super-hydrophilicity, Soft Matter, 5, 2009, 275-281.

3. F. Zhou, T. S. W. Huck, Surface grafted polymer brushes as ideal building blocks for "smart" surfaces, Physical Chemistry Chemical Physics, 8, 2006, 3815-3823.

4. G. Demirel, Z. Rzaev, S. Patir, E. Piskin, Poly(N-isopropylacrylamide) Layers on Silicon Wafers as Smart DNA-Sensor Platforms, Journal of Nanoscience and Nanotechnology, 9, 2009, 1865-1871.

5. F. Guo, Z. Guo, Inspired smart materials with external stimuli responsive wettability: A review, RSC Advances, 6, 2016, 36623-36641.

6. T. Sun, L. Feng, X. Gao, L. Jiang, Bioinspired surfaces with special wettability, Accounts of Chemical Research, Vol. 38, No. 8, 2005, 644-652.

7. Y. Liu, L. Mu, B. Liu, J. Kong, Controlled switchable surface, Chemistry A European Journal, 11, 2005, 2622-2631.

8. Z. Zheng, O. Azzaroni, F. Zhou, T. S. W. Huck, Topography printing to locally control wettability, Journal of American Chemistry Society, Vol. 128, 24, 2006, 7730-7731.

9. V. Krokos, G. Pashos, N. A. Spyropoulos, G. Kokkoris, G. A., Papathanasiou G. A. Boudouvis, Optimization of patterned surfaces for improved superhydrophobicity through cost-effective large-scale computations, Langmuir, 35, 2019, 6793-6802.

10. Y. C. Kuo, C. Gau, Control of superhydrophilicity and superhydrophobicity of a superwetting silicon nanowire surface, Journal of the Electrochemical Society, Vol. 157, 9, 2010, K201-K205.

11. Q. Yu, X. Li, Y. Zhang, L. Yan, T. Zhao, H. Chen, The synergistic effects of stimuli-responsive polymers with nano-structured surfaces: wettability and protein adsorption, RSC Advances, 1, 2011, 262-269.

12. Q. Fu, G. V. R. Rao, B. S. Basame, J. D. Keller, K. Artyushkova, E. J. Fulghum, P. G. Lopez, Reversible control of free energy and topography of nanostructured surfaces, Journal of American Chemistry Society, 126, 2004, 8904-8905.

13. T. Sun, G. Wang, L. Feng, B. Liu, Y. Ma, L. Jiang, D. Zhu, Reversible switching between superhydrophilicity and superhydrophobicity, Angewandte Chemie International Edition, 43, 2004, 357-360.

14. T. C. Schwall, A. I. Banerjee, Micro- and nanoscale hydrogel systems for drug delivery and tissue engineering, Materials, 2009, 2, 577-612.

15. L. Li, Y. Zhu, B. Li, C. Gao, Fabrication of thermoresponsive polymer gradients for study of cell adhesion and detachment, Langmuir, 24, 2008, 13632-13639.

16. D. M. Kurkuri, R. M. Nussio, A. Deslandes, H. N. Voelcker, Thermoresponsive copolymer coatings with enhanced wettability switching, Langmuir, 24, 2008, 4238-4244.

17. T. Sun, W. Song, L. Jiang, Control over the responsive wettability of poly(N-isopropylacrylamide) film in a large extent by introducing an irresponsive molecule, Chemical Communications, 2005, 1723-1725.

18. S. Kumar, L. Y. Dory, M. Lepage, Y. Zhao, Surface-Grafted Stimuli-Responsive Block Copolymer Brushes for the Thermo-, Photo- and pH-Sensitive Release of Dye Molecules, Macromolecules, 44, 2011, 7385-7393.

19. K. J. Chen, H. J. Wang, K. S. Fan, Chang Y. J., Reversible hydrophobic/hydrophilic adhesive of PS-b-PNIPAAm copolymer brush nanopillar arrays for mimicking the climbing aptitude of geckos, The Journal of Physical Chemistry C, 116, 2012, 6980-6992.

20. Q. Yu, Y. Zhang, H. Chen, F. Zhou, Z. Wu, H. Huang, L. J., Brash Protein adsorption and cell adhesion/detachment behavior on dual-responsive silicon surfaces modified with poly(N-isopropylacrylamide)-block-polystyrene copolymer, Langmuir, Vol. 26, 11, 2010, 8582-8588.

21. E. Stamm, Polymer Surfaces and Interfaces, Characterization, Modification and Applications, Springer, 2008.

22. Q. Zheng, C. LV, P. Hao, J. Sheridan, Small is beautiful, and dry, Science China, Physics, Mechanics &Astronomy, 53, 2010, 2245-2259.

23. D. Quere, Wetting and Roughness, Annual Review of Materials Research, 38, 2008, 71-79.

24. M. Callies, D. Quere, On water repellency, Soft Matter, 1, 55-61, 2005.

25. A. Pimpin, W. Srituravanich, Review on micro- and nanolithography techniques and their applications, Engineering Journal, Vol. 16, 1, 2012, 37-55.

26. F. M. Wang, N. Raghunathan, B. Ziaie, A non-lithographic top-down electrochemical approach for creating hierarchical (micro-nano) superhydrophobic silicon surfaces, Langmuir, 23, 2007, 2300-2303.

27. Y. Kwon, N. Patankar, J. Choi, J. Lee, Design of surface hierarchy for extreme hydrophobicity, Langmuir, Vol. 25, 11, 2009, 6129-6136.

28. Y. Xiu, L. Zhu, W. D. Hess, P. C. Wong, Hierarchical silicon etched structures for controlled hydrophobicity/superhydrophobicity, Nano Letters, Vol. 7, 11, 2007, 3388-3393.

29. Y. A. Vorobyev, C. Guo, Direct femtosecond laser surface nano/microstructuring and its applications, Laser Photonics Reviews, 7, No. 3, 2013, 385-407.

30. G. D. Kotsifaki, M. Kandyla, G. P. Lagoudakis, Near-field enhanced optical tweezers utilizing femtosecond-laser nanostructured substrates, Applied Physics Letters, 107, 2015, 211111.

31. R. R. Gattass, E. Mazur, Femtosecond laser micromachining in transparent materials, Nature Photonics, 2, 2008, 219-225.

32. M. Kanidi, A. Papagiannopoulos, A. Skandalis, M. Kandyla, S. Pispas, Thin films of PS/PS-b-PNIPAM and PS/PNIPAM polymer blends with tunable wettability, Journal of Polymer Science Part B: Polymer Physics, Vol. 57, 11, 2019, 670-679.

33. G. D. Kotsifaki, M. Kandyla, G. P. Lagoudakis, Plasmon enhanced optical tweezers with gold-coated black silicon, Scientific Reports, 6, 2016, 26275.

34. M. Kanidi, A. Dagkli, N. Kelaidis, D. Palles, S. Aminalragia-Giamini, J. Marquez-Velasco, A. Colli, A. Dimoulas, E. Lidorikis, M. Kandyla, I. E. Kamitsos, Surface-enhanced Raman spectroscopy of graphene integrated in plasmonic silicon platforms with a three-dimensional nanotopography, Journal of Physical Chemistry C, 123 (5), 2019, 3076-3087.

35. K. Norrman, A. Ghanbari-Siahkali, B. N. Larsen, 6 Studies of spin-coated polymer films, Annual Report Progress in Chemistry, Section C: Physical Chemistry, 101, 2005, 174-201.

36. H. C. Crouch, E. J. Carey, M. J. Warrender, J. M. Aziz, E. Mazur, Comparison of structure and properties of femtosecond and nanosecond laser-structures silicon, Applied Physics Letters, 84, 2004, 1820.

37. H. D. Lowndes, D. J. Fowlkes, J. A. Pedraza, Early stages of pulsed-laser growth of silicon microcolumns and microcones in air and $SF_6$, Applied Surface Science, 154-155, 2000, 647-658.

38. R. B. Tull, E. J. Carey, E. Mazur, P. J. McDonald, M. S. Yalisove, Silicon surface morphologies after femtosecond laser irradiation, MRS Bulletin, 31, 2006, 626-633.

39. G. Amoako, Femtosecond laser structuring of materials: a review, Applied Physics Research, Vol. 11, 3, 2019, 1-11.

40. T. Y. Wu, A. B. Zrimsek, S. V. Bykov, R. S. Jakubek, S. A. Asher, Hydrophobic collapse initiates the poly(N-isopropylacrylamide) volume phase transition reaction coordinate, Journal of Physical Chemistry B, 122, 2018, 3008-3014.

41. M. Xia, S. Pan, Y. Sun, Q. Luo, Z. Fang, P. He, J. Fu, Y. Zhang, Dehydration behaviours of isopropyl group initiate the surface wettability transition of temperature sensitive poly(N-isopropylacrylamide) hydrogel, Material Research Express, 6, 2019, 095704.

42. J. Dybal, M. Trchova, P. Schmidt, The role of water in structural changes of poly(N-isopropylacrylamide) and poly(N-isopropylmethacrylamide) studied by FTIR, Raman spectroscopy and quantum chemical calculation, Vibrational Spectroscopy, 51, 2009, 44-51.

43. S. F. Bates, Polymer-polymer phase behavior, Science, 251, 1991, 898-905.

44. Y. He, C. Jiang, H. Yin, W. Yuan, Tailoring the wettability of patterned silicon surfaces with dual-scale pillars: From hydrophilicity to superhydrophobicity, Applied Surface Science, 257, 2011, 7689-7692.

**Figure captions**

**Figure 1:** Scanning electron micrographs (SEM) of microstructured silicon surfaces (a) with a native $SiO_2$ layer and (b) without it.

**Figure 2:** (a) Scanning electron micrograph (SEM) of a 50/50 PS/PNIPAM film casted on a microstructured silicon substrate (without $SiO_2$). (b) Higher magnification.

**Figure 3:** (a) Layered EDS image of a 50/50 PS/PNIPAM film casted on a microstructured silicon substrate (without $SiO_2$) and (b) mapping distribution of elements.

**Figure 4:** Raman spectra, measured with 514 nm excitation wavelength, of (a) a 75/25 PS/PNIPAM film casted on a microstructured silicon substrate (with $SiO_2$) and (b) a 50/50 PS/PNIPAM film casted on a microstructured silicon substrate (with $SiO_2$). For clarity, Raman spectra have been vertically shifted in Fig. 4a. Each spectrum in Figs. 4a and b corresponds to a different location on the film surface.

**Figure 5:** Water contact angle measurements on flat and microstructured silicon surfaces and on films of PS/PNIPAM blends, casted on flat and microstructured silicon substrates with or without a native $SiO_2$ layer, at $25^oC$.

**Figure 6:** Water contact angle measurements on PS/PNIPAM films casted on flat and microstructured silicon substrates (a) with a native $SiO_2$ layer and (b) without it, at $25^oC$ and $45^oC$.

**Figure 7:** (a) Image of a water droplet on a 50/50 PS/PNIPAM film, casted on a microstructured silicon substrate with a native $SiO_2$ layer, at $25^oC$ and $45^oC$. (b) Water contact angle measurements for a similar film/substrate combination for seven cycles of heating/cooling at $45^oC$ and $25^oC$. Dashed lines are guides to the eye for the trend of the contact angles at $25^oC$ (blue color) and $45^oC$ (red color).

**Figure 8:** Elemental quantification (%) of a 50/50 PS/PNIPAM film casted on flat silicon (a) with a native $SiO_2$ layer and (b) without it, as a function of number of etching layers, using X-ray photoelectron spectroscopy (XPS).

**Figure 9:** (a) Ratio of PNIPAM, calculated by the variation of the relative amount of N-C=O population and (b) ratio of PS, calculated by the relative amount of the shake-up $\pi$-$\pi$* transition population, in a 50/50 PS/PNIPAM film casted on flat silicon with a native $SiO_2$ layer. (c) Ratio of PNIPAM, calculated by the variation of the relative amount of N-

C=O population and (d) ratio of PS, calculated by the variation of the relative amount of the shake-up $\pi$-$\pi$* transition population, in a 50/50 PS/PNIPAM film casted on flat silicon without a native $SiO_2$ layer. Details about the calculations are presented in supplementary material. Red lines indicate least squares fit to the data.

Figure 1

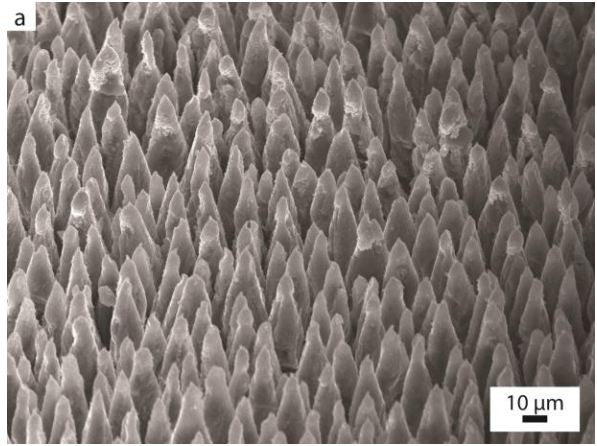

Figure 1a

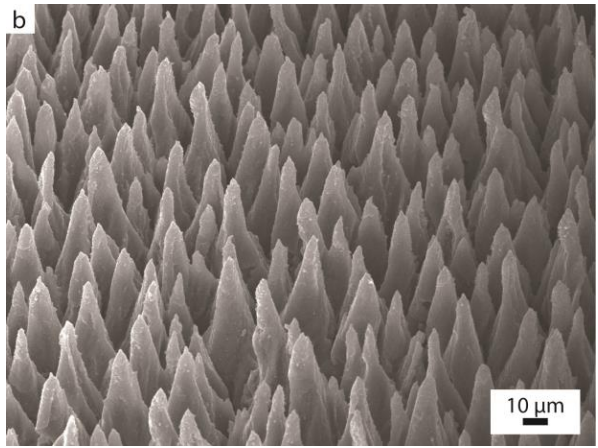

Figure 1b

Figure 2

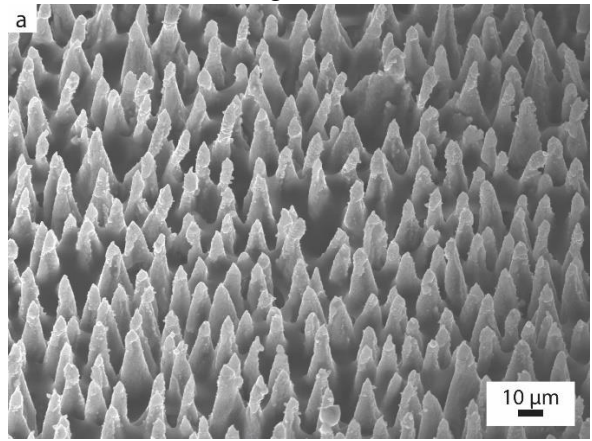

Figure 2a

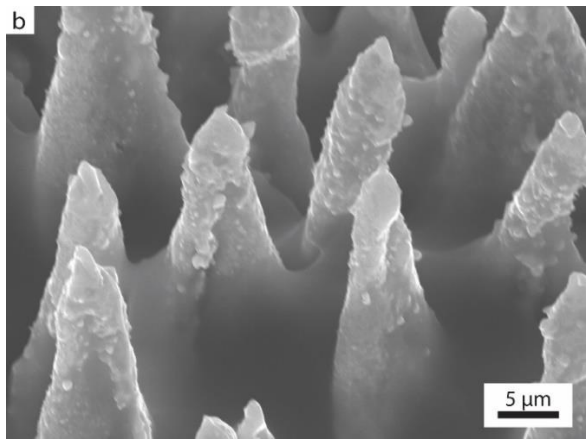

Figure 2

Figure 3

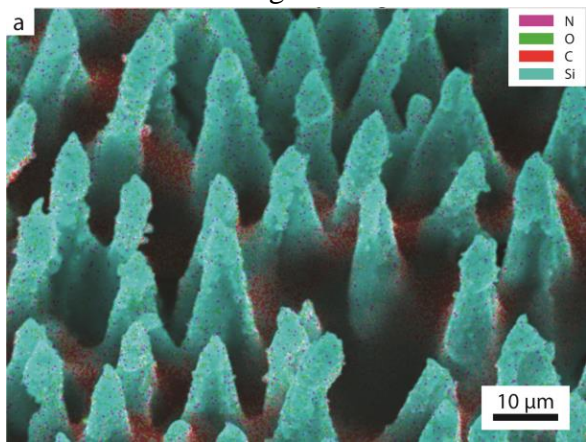

Figure 3a

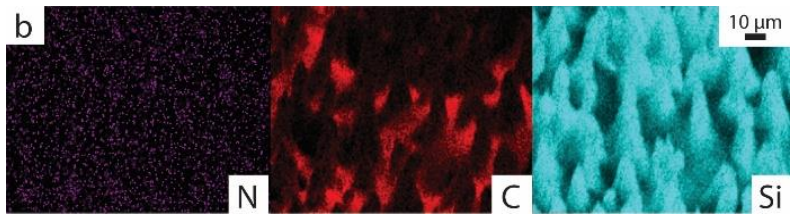

Figure 3b

Figure 4

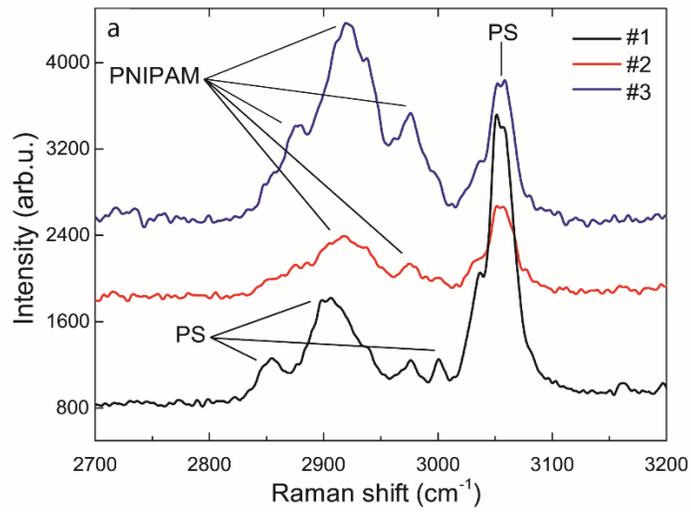

Figure 4a

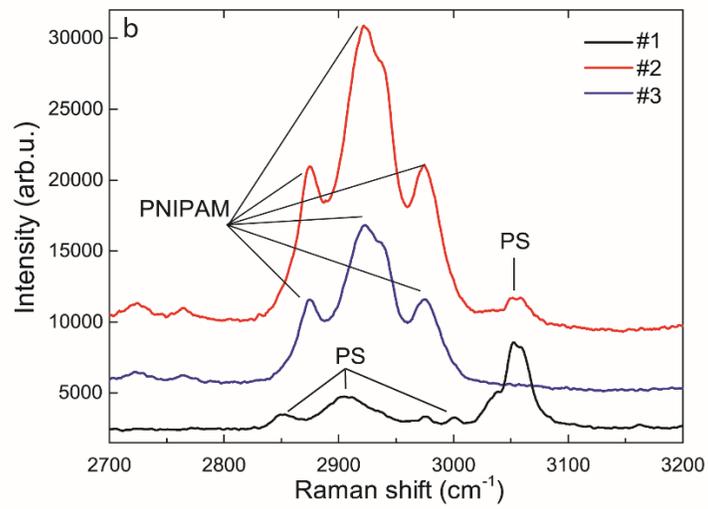

Figure 4b

Figure 5

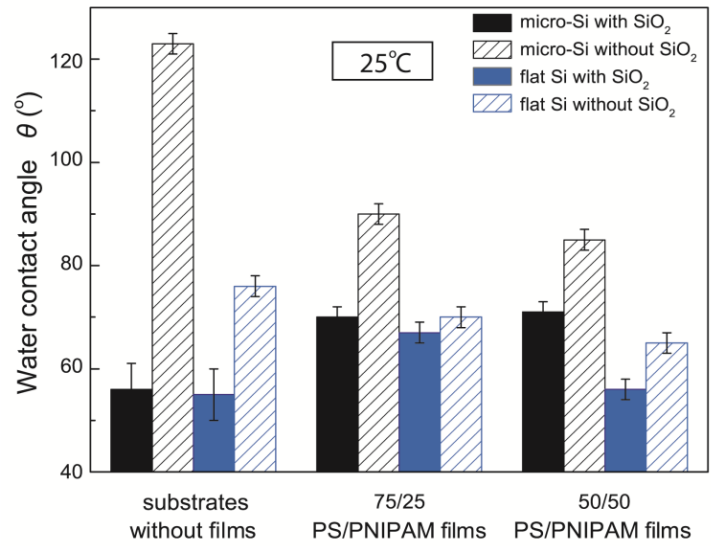

Figure 6

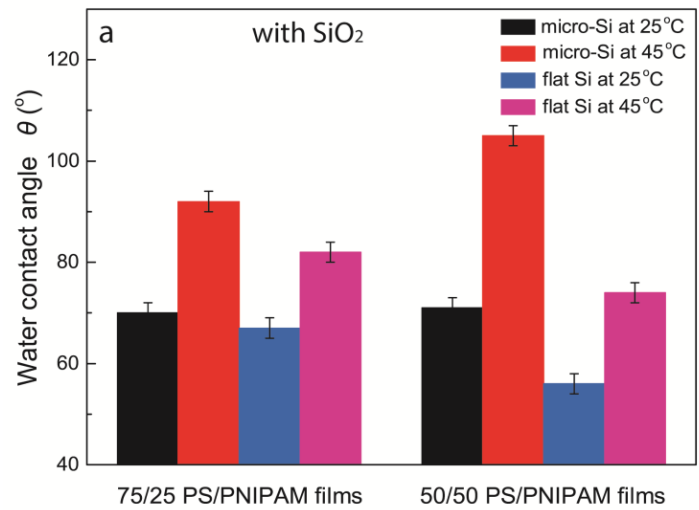

Figure 6a

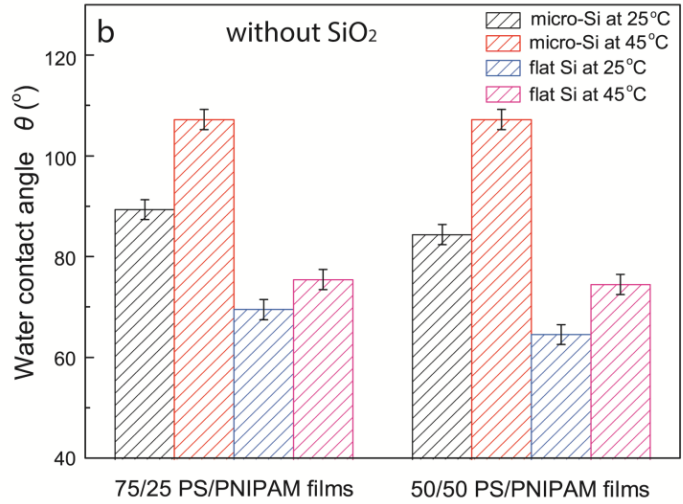

Figure 6b

Figure 7

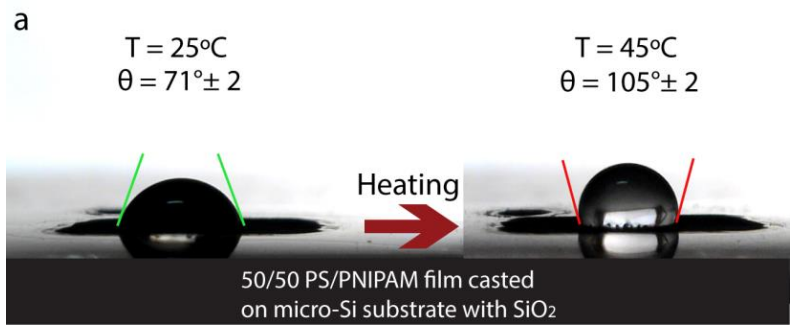

Figure 7a

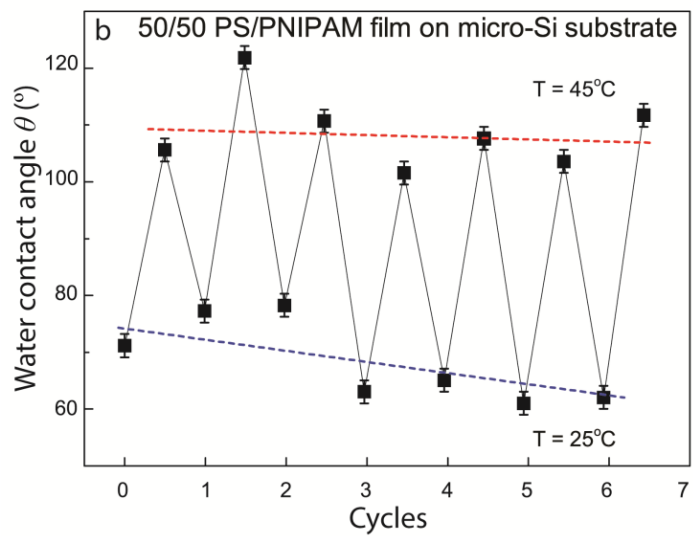

Figure 7b

Figure 8

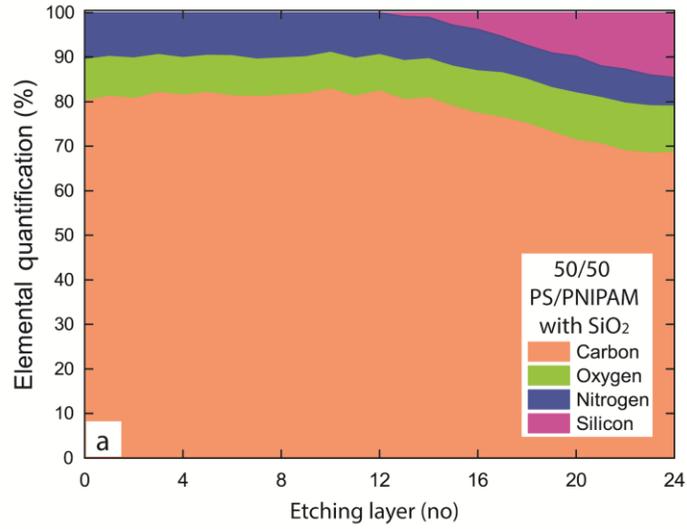

Figure 8a

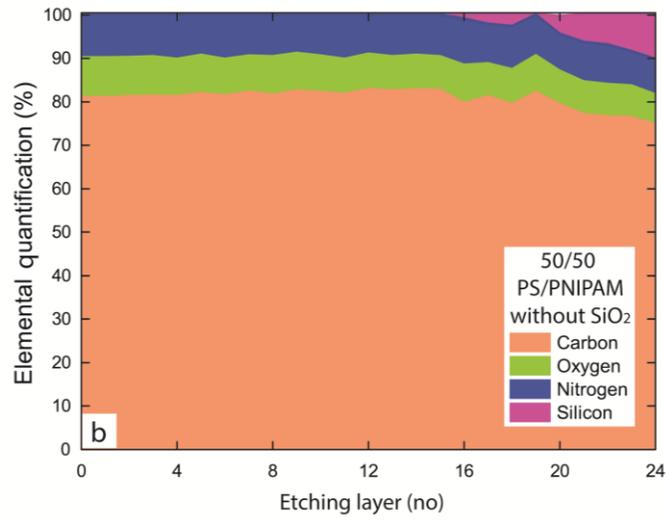

Figure 8b

Figure 9

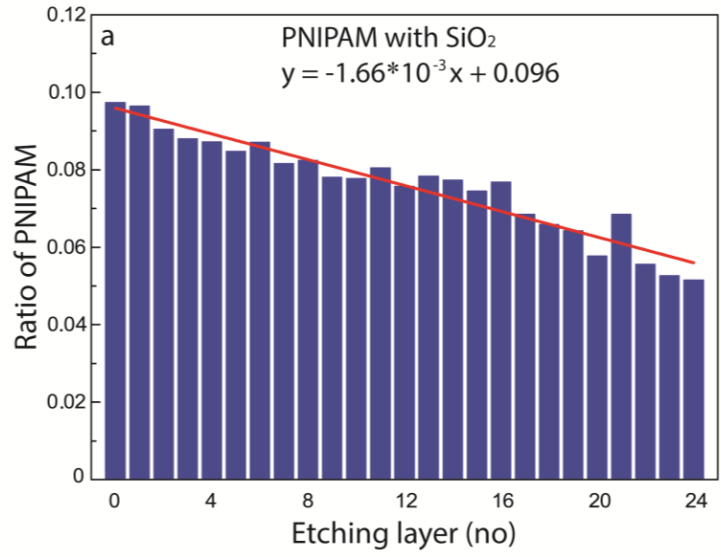

Figure 9a

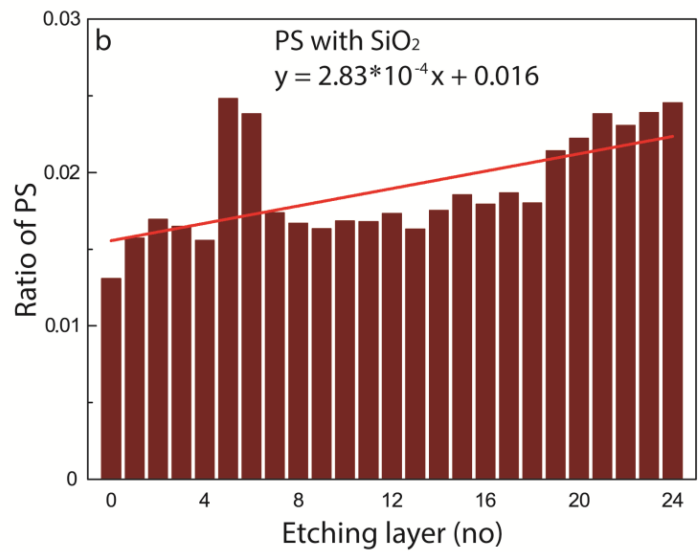

Figure 9b

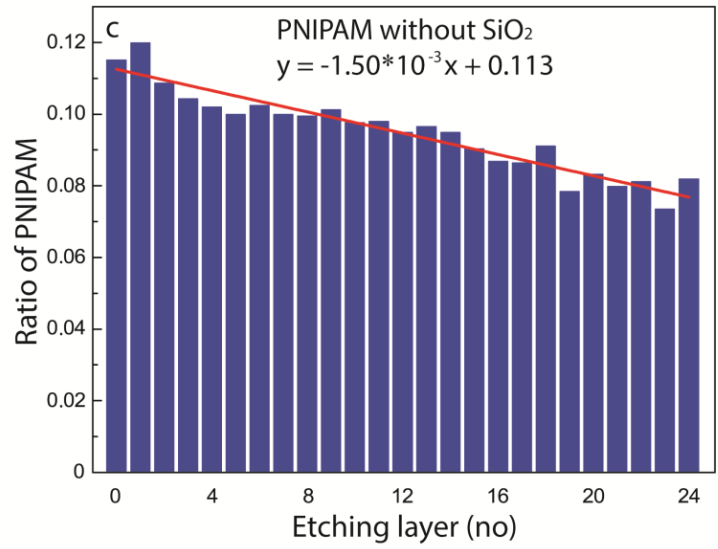

Figure 9c

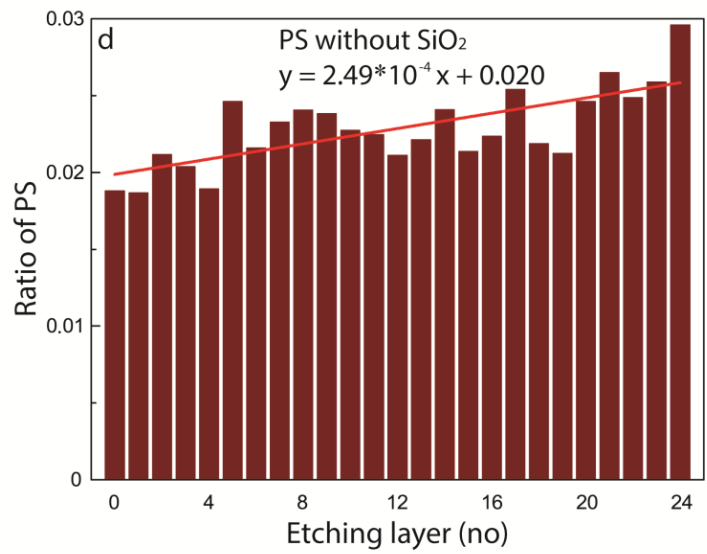

Figure 9d